\documentclass[aps, prl, 10pt, twocolumn, superscriptaddress, noshowpacs, preprintnumbers, floatfix]{revtex4-2}

\usepackage[dvipsnames]{xcolor}
\usepackage{mathtools}
\usepackage{amsmath}
\usepackage{amsfonts}
\usepackage{amssymb}
\usepackage{bbold}
\usepackage{upgreek}
\usepackage{multirow}
\usepackage{bm}
\long\def\exclude#1{}
\usepackage{orcidlink}
\usepackage{slashed}
\usepackage{url}

\usepackage{graphicx}
\usepackage{color}

\begin{document}

\title{GeV-scale QCD Axion}

\author{Hitoshi Murayama}
\affiliation{Leinweber Institute for Theoretical Physics, University of California, Berkeley, CA 94720, USA}
\affiliation{Kavli Institute for the Physics and Mathematics of the Universe (WPI), University of Tokyo, Kashiwa 277-8583, Japan}
\affiliation{Ernest Orlando Lawrence Berkeley National Laboratory, Berkeley, CA 94720, USA}

\begin{abstract}
In order to solve the strong CP problem, we study the possibility that the Peccei--Quinn symmetry is broken {\it below}\/ the QCD scale. We find that a QCD axion can be above GeV, and may be among the observed $\eta$ resonances. It is immune to quantum gravity corrections. The only fermion that has a $U(1)$ Peccei--Quinn charge is the right-handed up quark. Flavor-changing neutral currents are surprisingly small. All accelerator and astrophysical limits can be evaded.  
The most significant constraint is the mass splitting between $\pi^\pm$ and $\pi^0$. In a UV completed model, LHC can look for a heavy quark pair ${\cal D} \overline{\cal D}$ followed by the decay ${\cal D} \rightarrow W+u$ or a single production $q u \rightarrow q\, {\cal U}$ followed by ${\cal  U} \rightarrow u Z, uh$. There can be an $O(1)$ contribution $h \rightarrow u\bar{u}\varphi$ in the measurement of $h \rightarrow gg$ or permille effects on the hadronic $Z$ width at a Higgs factory. 
\end{abstract}

\maketitle

\section{Introduction}

The strong CP problem is considered to be one of the major mysteries in the standard model of particle physics  (see {\it e.g.}\/, \cite{Hook:2018dlk} for a recent review). The theory of strong interaction, Quantum ChromoDynamics (QCD), can have a topological term in its Lagrangian
\begin{align}
	{\cal L}_\theta &=\frac{\theta}{16\pi^2} {\rm Tr}\, G_{\mu\nu}\tilde{G}^{\mu\nu}\ ,
\end{align}
which can break CP and time-reversal symmetries. The parameter $\theta$ is called the vacuum angle given its periodicity by $2\pi$. Its presence would induce a non-zero neutron electric dipole moment
\begin{align}
	H = -d_n \vec{s}_n \cdot \vec{E}
\end{align}
with an estimate $d_n =  1.48(14)(31) \times 10^{-16} \theta~e~{\rm cm}$ from a recent lattice QCD calculation \cite{Liang:2023jfj}. Given the experimental upper limit $d_n < 1.8 \times 10^{-26}~e~{\rm cm}$ \cite{Abel:2020pzs}, the angle $\theta$ must be exceedingly small $\theta \lesssim 10^{-10}$ while theoretically it can be as large as $\pi$ in principle. To understand why an angle is so small is the strong CP problem (see, {\it e.g.}\/, \cite{benabou2025clearingstrongcpproblem} for a recent discussion why this is indeed a problem).

A new particle called axion has been regarded as a leading candidate to explain the strong CP problem. The idea is to have a global $U(1)_{PQ}$ symmetry in the classical Lagrangian which is spontaneously broken, and the resulting Nambu--Goldstone boson is the axion. If the $U(1)_{PQ}$ symmetry is anomalous under QCD, the axion effectively replaces the $\theta$ parameter as
\begin{align}
	\theta \rightarrow \theta + \frac{a}{f_a}\ ,
\end{align}
where $f_a$ is  the energy scale where the $U(1)_{PQ}$ symmetry is broken and is called the axion decay constant \cite{Peccei:1977hh,Peccei:1977ur}. The axion settles to the minimum of the potential where $\theta$ is dynamically cancelled with $\langle a \rangle = - f_a \theta$. However the original proposal by Peccei and Quinn is experimentally excluded by beam dump experiments and search for $J/\psi, \Upsilon \rightarrow a \gamma$ etc as it assumed $f_a \approx v$ where $v = \langle H^0 \rangle = 174$~GeV is the energy scale of the electroweak symmetry breaking. 

``Invisible'' axion models were proposed to make the axion hypothesis phenomenologically viable by assuming $f_a \gg v$, Since the axion interaction is suppressed by $1/f_a$, a large $f_a$ can evade experimental constraints. The concrete models are broadly classified as DFSZ axion models \cite{Dine:1981rt,Zhitnitsky:1980tq} if the standard model quarks and leptons have $U(1)_{PQ}$ charges, or KSVZ axion models \cite{Kim:1979if,Shifman:1979if} if $U(1)_{PQ}$ charges are carried by new colored fermions. Astrophysical constraints typically place a very high lower limit on $f_a$. For instance, the proto-neutron star core of SN1987a or neutron stars would cool too fast if $f_a \lesssim 5 \times 10^8$~GeV \cite{Raffelt:2006cw,Hamaguchi:2018oqw}. For $f_a \gtrsim 10^{11}$~GeV, the misalignment production or cosmic strings can produce axion as a viable dark matter candidate \cite{Bae:2008ue,Benabou:2023npn}. This is an attractive hypothesis that hits two birds (strong CP problem and dark matter) with a single stone (axion). There is an active effort to search for a dark-matter axion experimentally (see {\it e.g.}\/ \cite{Giannotti:2024xhx} for a recent review).

Unfortunately, such ``invisible'' axion models are subject to potentially disastrous quantum-gravity corrections \cite{Kamionkowski:1992mf,Holman:1992us,Barr:1992qq}. It is often estimated that the $U(1)_{PQ}$ global symmetry is explicitly broken by powers in $f_a / M_{Pl}$, and a high $f_a$ makes the theory more susceptible to such effects. One way to avoid this problem is to obtain $U(1)_{PQ}$ as an accidental low-energy symmetry for a composite axion \cite{Choi:1985cb,Randall:1992ut} (see {\it e.g.}\/, \cite{Gherghetta:2025fip,gherghetta2025highqualityaxionexactsusy} for recent concrete models).

In this paper, we explore the possibility to go the opposite direction by assuming $f_a$ is very low $f_a \ll v$, in fact below the QCD scale. At the first sight, such a low $f_a$ seems to be immediately excluded by a myriad of experimental constraints because the axion coupling becomes stronger than the original proposal by Peccei and Quinn. Surprisingly, we find that the experimental and astrophysical constraints are not necessarily fatal. The most stringent limit comes from the mass splitting between charged and neutral pions. The axion and its scalar partner are in the 1--2~GeV range and may well be among the scalar $f_0$ and pseudo-scalar $\eta$ resonances in the hadronic spectroscopy. Given the low $f_a$, it is immune to quantum-gravity corrections.

\section{The Main Idea}

The main idea is to have the $U(1)_{PQ}$ symmetry at the QCD scale. It is implemented with a complex scalar $\varphi$ with the PQ charge $+1$ and the right-handed up quark $u_R$ with the PQ charge $-1$. No other particles are charged under $U(1)_{PQ}$ at the QCD scale.

The Lagrangian at the QCD scale is
\begin{align}
	{\cal L}_{QCD} =& - \frac{1}{2} {\rm Tr}\, G_{\mu\nu}G^{\mu\nu}
	+ \frac{\theta}{16\pi^2} {\rm Tr}\, G_{\mu\nu}\tilde{G}^{\mu\nu}
	- \frac{1}{4} F_{\mu\nu} F^{\mu\nu} 
	\nonumber \\
	& + \bar{u} i{\not\!\! D} u + \bar{d} (i{\not\!\! D}-m_d) d + \bar{s} (i{\not\!\! D}-m_s) s
	\nonumber \\
	& + \partial_\mu \varphi^* \partial^\mu \varphi  - m_\varphi^2 \varphi^* \varphi
	- (\kappa \varphi\, \overline{u_L} u_R + c.c.).
	\label{eq:QCD}
\end{align}
Here, $\kappa$ is a Yukawa coupling of $\varphi$ to the up quark. The Lagrangian is classically invariant under $U(1)_{PQ}$ but it is anomalous under QCD because of the rotation of $u_R$. As a result, we can remove $\theta$ by the $U(1)_{PQ}$ rotation from the Fujikawa measure of the fermion path integral \cite{Fujikawa:1979ay,Fujikawa:1980eg}, $\varphi\rightarrow e^{-i\theta}\varphi$, $u_R \rightarrow e^{i\theta} u_R$, solving the strong CP problem.

\section{Up Quark Mass}

There is no mass term for the up quark in Eq.~\eqref{eq:QCD} because it is forbidden by the $U(1)_{PQ}$ symmetry. However the chiral symmetry breaking in QCD induces a linear term in $\varphi$ which makes it acquire a vacuum expectation value producing the up-quark mass. This mechanism is akin to the bosonic technicolor theory of electroweak symmetry breaking with elementary Higgs doublet \cite{Samuel:1990dq}.

Switching to the chiral Lagrangian with $U = e^{2i \pi^a T^a/f_\pi}$, we obtain
\begin{align}
	{\cal L}_{\chi} =& \frac{f_\pi^2}{4} {\rm Tr} \partial_\mu U^\dagger \partial^\mu U
	+ \partial_\mu \varphi^* \partial^\mu \varphi - m_\varphi^2 \varphi^* \varphi
	\nonumber \\
	& + \mu^3 {\rm Tr} M U + c.c.,
	\label{eq:chiral}
\end{align}
where $\mu^3 = f_\pi^2 m_\pi^2 / 2(m_u + m_d) \approx 0.022~{\rm GeV}^3$, with the quark mass matrix given by
\begin{align}
	M =& \left( \begin{array}{ccc} \kappa \varphi & &\\
		& m_d &\\ & & m_s \end{array} \right) .
\end{align}
Here and below we adopt the PDG averages of quark masses defined in the $\overline{\rm MS}$ scheme at the renormalization scale of 2~GeV as reference values \cite{ParticleDataGroup:2024cfk}. Recall that the matrix element $\mu^3 = \langle \overline{u}_L u_R \rangle$ also runs in such a way that the product $m_q \mu^3$ is renormalization-group invariant. Due to the linear term in $\varphi$ it acquires an expectation value 
\begin{align}
	\langle \varphi \rangle &= \frac{\kappa \mu^3}{m_\varphi^2}\ ,
	\label{eq:varphi}
\end{align}
which in turn generates the up-quark mass
\begin{align}
	m_u =& \kappa \langle \varphi \rangle =  \frac{\kappa^2 \mu^3}{m_\varphi^2}\, .
\end{align}
Note that the lattice determination of quark masses averaged in \cite{ParticleDataGroup:2024cfk} assume the standard model, and the presence of the pion-axion mixing discussed below may change the values. For consistency, we need to allow for an effective up-quark mass subject to the Kaplan--Manohar ambiguity \cite{Kaplan:1986ru}. With the observation that the quark mass matrix $M={\rm diag}(m_u,m_d,m_s)$ and the combination ${\rm det}M^* (M^\dagger)^{-1}$ have the same chiral transformation property, the up-quark mass can have an additional contribution of $O(m_d m_s/4\pi f_\pi)$ which is sizable. When we use the PDG value of the up-quark mass $m_u = 2.2$~MeV, we take the point of view that it applies to the sum of these two contribution, 
\begin{align}
	m_u^{\it eff} = m_u + O(1) \frac{m_d m_s}{4\pi f_\pi} = R m_u,
\end{align}
where $R$ is the enhancement factor. This ambiguity has been used in the past to argue that the up-quark mass $m_u$ may actually be zero, making the strong CP problem non-existent. That is not what we attempt here. We rather allow for an enhancement of the order of a few. Given this enhancement, and to obtain the required total mass $m_u^{\it eff} = 2.16 \pm 0.04$~MeV, we need
\begin{align}
	m_\varphi &= \left( \frac{\mu^3 \kappa^2}{m_u^0} \right)^{1/2} = 3.2~{\rm GeV} \kappa R^{1/2}.
	\label{eq:mphi}
\end{align}
As we will see below, we need $\kappa \lesssim 0.5$ in a peturbative UV completion of this model and hence the mass needs to be lighter than about 2~GeV. 

\section{Quantum Gravity Corrections}

A potential leading quantum gravity correction to the axion potential would be
\begin{align}
	V \supset c \frac{\varphi^5}{M_{Pl}} + c.c.\, ,
\end{align}
with $c$ being a complex number with $|c| \sim O(1)$. A potential shift in the minimum would be
\begin{align}
	\Delta \theta \approx \Im m(c) \frac{1}{m_\varphi^2} \frac{\langle\varphi\rangle^3}{M_{Pl}} 
	= 1.3 \times 10^{-27} \frac{\rm GeV^2}{\kappa^3 m_\varphi^2} \Im m(c )  .
\end{align}
For $m_\varphi$ in the GeV range and $\kappa \simeq 0.5$, the shift is completely negligible compared to the experimental limit $\theta \lesssim 10^{-10}$, fulfilling our goal.

\section{The Axion Mass}

To study phenomenology, we need to look at the mass spectrum. With the Lagrangian Eq.~\eqref{eq:QCD}, the complex scalar
\begin{align}
	\varphi = \frac{1}{\sqrt{2}} (\sigma + i a)
\end{align}
has a degenerate pair of an axion $a$ and its scalar partner $\sigma$ of mass approximately $m_\varphi$. If we include a $(\varphi^* \varphi)^2$ term in the potential, we can break the degeneracy and the axion becomes lighter than the scalar. We do not need this coupling and ignore it for much of the discussions below, unless it qualitatively modifies the result. It is important that the axion mass does not vanish even when $m_u \rightarrow 0$ or $\kappa \rightarrow 0$ unlike the conventional axion models.

Note that what we call an axion here is defined as ``a new particle that appears in theories that solve the strong CP problem,'' not as ``a pseudo-Nambu--Goldstone boson that acquires a mass from the QCD chiral anomaly.'' 

It may be indeed puzzling that the axion mass is determined by an external parameter $m_\varphi^2$ totally unrelated to the QCD scale. Normally we expect the axion mass to vanish in the absence of non-perturbative QCD dynamics and hence should be proportional to powers in $\Lambda_{\rm QCD}$, normally the second power $m_a \propto \Lambda_{\rm QCD}^2/f_a$. It is not the case here. On the other hand if we turn off the QCD anomaly for $U(1)_A$, there is an exact Peccei--Quinn symmetry that is spontaneously broken and hence there must be a massless Nambu--Goldstone boson. 

To see all this, we need to study the mass matrix of neutral pseudoscalar states including the non-perturbative QCD effects. We use the standard basis including the $U(1)_A$ Nambu--Goldstone boson $\eta_1$, which is assumed to have a mass term $\frac{1}{2} m_1^2 \eta_1^2$ from the non-perturbative QCD effects (see \cite{Witten:1979vv,Veneziano:1979ec} for large $N_c$ arguments, and \cite{Csaki:2023yas,Kondo:2025njf} for analytic results using supersymmetry), 
\begin{widetext}
\begin{align}
	\pi^a T^a = \left( \begin{array}{ccc} 
	\frac{1}{2} \pi_3 + \frac{1}{2\sqrt{3}}\eta_8 + \frac{1}{\sqrt{6}} \eta_1 & \frac{1}{\sqrt{2}} \pi^+ & \frac{1}{\sqrt{2}} K^+ \\
	\frac{1}{\sqrt{2}} \pi^- & -\frac{1}{2} \pi_3 + \frac{1}{2\sqrt{3}}\eta_8 + \frac{1}{\sqrt{6}} \eta_1 &  \frac{1}{\sqrt{2}} K^0 \\
	\frac{1}{\sqrt{2}} K^- & \frac{1}{\sqrt{2}} \overline{K}^0 & - \frac{1}{\sqrt{3}} \eta_8+ \frac{1}{\sqrt{6}} \eta_1 
	\end{array} \right).
\end{align}
Then the pseudoscalars have the mass matrix
\begin{align}
	V \supset \frac{1}{2} ( \begin{array}{cccc}a & \pi_3 & \eta_8 & \eta_1 \end{array} )
\left(
\begin{array}{cccc}
	m_\varphi^2 
	& \sqrt{2} \frac{\kappa  \mu^3}{f_\pi} 
	& \sqrt{\frac{2}{3}} \frac{\kappa  \mu^3}{f_\pi} 
	& \frac{2}{\sqrt{3}}\frac{\kappa \mu^3}{f_\pi} \\
	\sqrt{2} \frac{\kappa  \mu^3}{f_\pi} 
	& 2 \frac{\mu^3}{f_\pi^2}  (m_u+m_d)
	& \frac{2}{\sqrt{3}} \frac{\mu^3}{ f_\pi^2 }  (m_u -m_d)
	& 2 \sqrt{\frac{2}{3}} \frac{\mu^3}{f_\pi^2}  (m_u -m_d)\\
	\sqrt{\frac{2}{3}}\frac{ \kappa  \mu^3}{f_\pi} 
	& \frac{2}{\sqrt{3}} \frac{\mu^3 }{f_\pi^2 } (m_u-m_d)
	& \frac{2}{3} \frac{\mu^3}{f_\pi^2}  (m_u+m_d+4 m_s )
	& \frac{2 \sqrt{2}}{3} \frac{\mu^3}{f_\pi^2} (m_u+m_d-2 m_s ) \\
	 \frac{2}{\sqrt{3}} \frac{\kappa  \mu^3}{f_\pi} 
	 & 2 \sqrt{\frac{2}{3}}\frac{ \mu^3 }{f_\pi^2 } (m_u-m_d )
	 & \frac{2 \sqrt{2}}{3} \frac{\mu^3}{f_\pi^2}  (m_u+m_d-2 m_s)
	 & \frac{4 \mu^3}{3 f_\pi^2}  (m_u+m_d+m_s)+m_1^2
\end{array}
\right) 
	\left( \begin{array}{c}a \\ \pi_3 \\ \eta_8 \\ \eta_1 \end{array} \right).
\label{eq:fullmass2}
\end{align}
\end{widetext}
The determinant of this matrix is
\begin{align}
	\frac{16}{3f_\pi^4}  \mu^6 m_d m_s m_1^2 m_\varphi^2 .
\end{align}
In the absence of the chiral anomaly, $m_1^2 = 0$ and indeed there is a massless eigenstate. The axion mass remains approximately $m_\varphi$ simply because it is the heaviest eigenstate in this mass matrix.

\section{Pion Mass Constraint}

The mass matrix Eq.~\eqref{eq:fullmass2} suggests that the mixing between the axion and the neutral pion pushes the mass of the neutral pion down by $\Delta m_{\pi^0}^2 = - 2\kappa^2 \mu^6 / f_\pi^2 m_\varphi^2 = - 2 \mu^3 m_u / f_\pi^2$ relative to that of the charged pion. To the leading order in $1/m_\varphi$, we find
\begin{align}
	m_{\pi^\pm}^2 &= \frac{1}{f_\pi^2} 2 \mu^3 (R m_u + m_d), \\
	m_{\pi^0}^2 &= \frac{1}{f_\pi^2} 2 \mu^3 ((R-1) m_u + m_d).
\end{align}
Normally we interpret the observed difference between the charged pion mass $m_{\pi^\pm} = 139.6$~MeV and the neutral pion mass $m_{\pi^0} = 135.0$~MeV as the electromagnetic correction, as studied by the lattice calculation, {\it e.g.}\/, in \cite{Fodor:2016bgu,Feng:2021zek}. Here we conservatively require that the mass shift to be at most the observed $m_{\pi^\pm}$--$m_{\pi^0}$ splitting of 4.6~MeV. We need
\begin{align}
	m_\varphi \gtrsim 6.7 \kappa~{\rm GeV}.
\end{align}
This limit is in a tension but can be reconciled with the requirement Eq.~\eqref{eq:mphi} by the enhancement factor $R$, and is the most significant constraint on our proposal \footnote{We thank Tsutomu Yanagida to point out this constraint.}. To settle this question, we need lattice calculations combined with chiral perturbation theory allowing for such a mixing in extracting quark masses from pseudoscalar and baryon octet masses. Note also that a higher-order chiral Lagrangian terms such as $\mu_2^2 {\rm Tr} M U M U$ can raise $m_\pi^0$ relative to $m_{\pi^\pm}$ even though it is usually assumed small $\mu_2^2 \approx \mu^3 / 4\pi f_\pi$ based on Naive Dimensional Analysis \cite{Georgi:1992dw}.

\section{Axion Decay}

Axion decays rapidly to $u\bar{u}$,
\begin{align}
	\Gamma(a \rightarrow u\bar{u})
	&= \frac{3}{16\pi} \kappa^2 m_a = 0.060 \kappa^2 m_a,
\end{align}
if computed perturbatively, or into pions,
\begin{align}
	\Gamma(a \rightarrow 3\pi^0)
	&= \frac{3}{128\pi^3} \frac{\kappa^2 \mu^6}{f_\pi^6} m_a
	= 0.080 \kappa^2 m_a, \\
	\Gamma(a \rightarrow \pi^0 \pi^+ \pi^-)
	&= \frac{1}{64\pi^3} \frac{\kappa^2 \mu^6}{f_\pi^6} m_a
	= 0.055 \kappa^2 m_a,
\end{align}
if computed with the chiral Lagrangian. They cannot be trusted, but provide rough estimates. Depending on its mass, the axion may decay also into $\pi^0 \pi^\pm K^\mp$, $\eta \pi^0 \pi^0$, $\eta \pi^+ \pi^-$, $\pi^0 K^+ K^-$, $\pi^0 \eta \eta$, $3\eta$. Clearly the axion is not dark matter because it decays with a lifetime of the order of $10^{-23}$~sec. The usual ``light-shining-through-a-wall'' search does not work due to its short lifetime.

\begin{figure}[t]
\includegraphics[width=0.8\columnwidth]{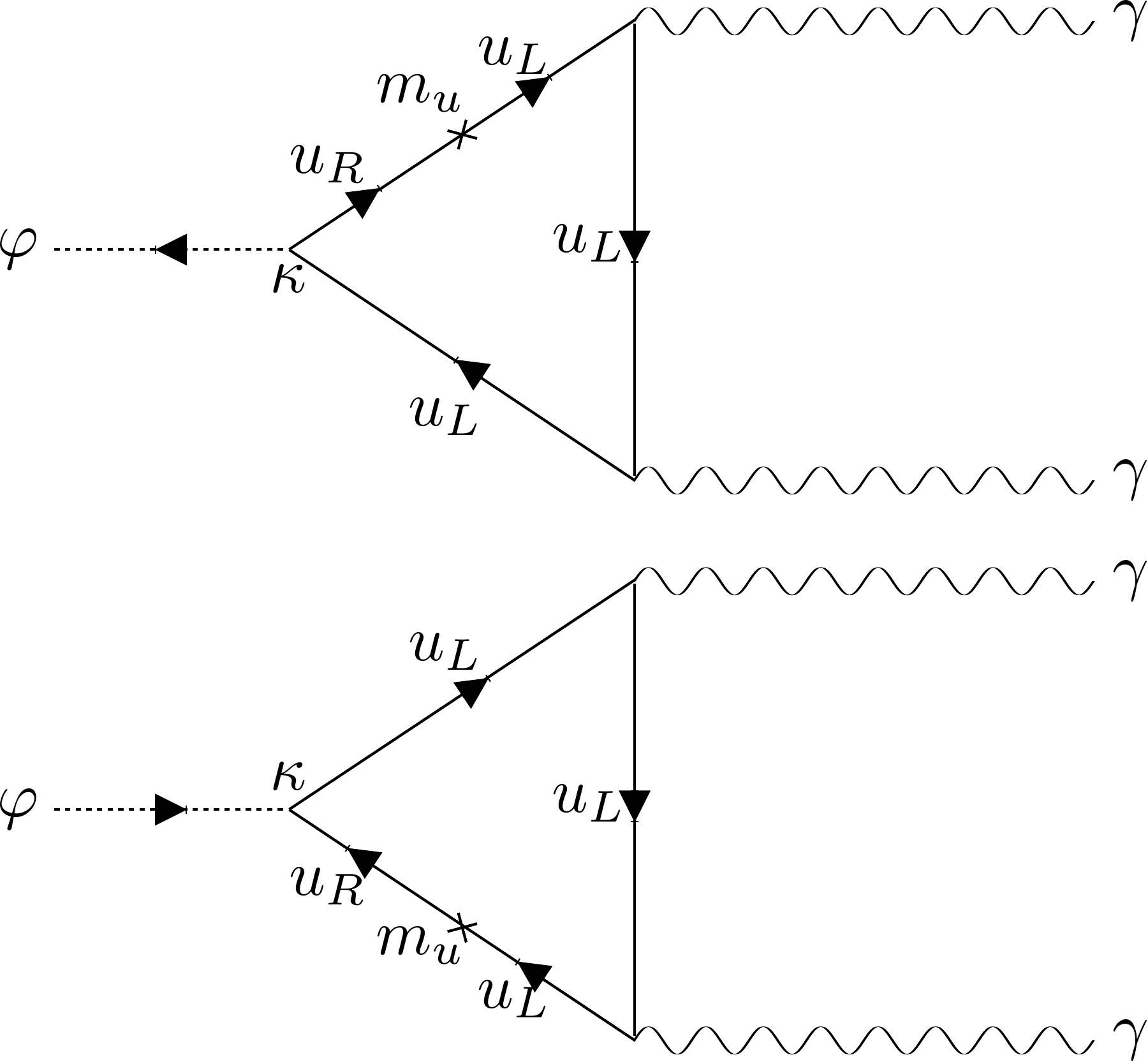}
\caption{The $aF\tilde{F}$ and $\sigma F F$ couplings induced by the up-quark loop. There are also diagrams where $u_R$ is coupled to photons, and diagrams with two photons exchanged need to be added. The mass insertion can appear on any of the three up-quark propagators. When photons are on-shell and $\varphi$ on-shell, it is suppressed as $\alpha \kappa m_u / 4\pi m_\varphi^2$ instead of $\alpha/f_a$. When photons are far off-shell $q^2 \neq 0$, including the case of the on-shell $Z$ boson $q^2 = m_Z^2$, the diagram is suppressed as $\alpha \kappa m_u / 4\pi q^2$ instead of $\alpha/f_a$. }
\label{fig:agammagamma}
\end{figure}

Compared to these hadronic final states, most experiments searched for $a \rightarrow e^+ e^-$, which is absent in our case, or $a\rightarrow\gamma\gamma$, negligible in our case. Indeed, the axion decay into $\gamma\gamma$ is suppressed by $(\alpha\kappa m_u/4\pi m_\varphi^2)^2$ as seen in Fig.~\ref{fig:agammagamma},  which is very small given $m_u \ll m_\varphi$. 

\section{Nuclear Force}

The coupling of $\varphi$ to the nucleons can be estimated as follows. The Gell-Mann--Okubo mass formula can be obtained from the effective Lagrangian of octet baryons 
\begin{align}
	B =& \left( \begin{array}{ccc} 
	\frac{1}{2} \Sigma^0 + \frac{1}{2\sqrt{3}}\Lambda^0 & \frac{1}{\sqrt{2}} \Sigma^+ & \frac{1}{\sqrt{2}} p \\
	\frac{1}{\sqrt{2}} \Sigma^- & -\frac{1}{2} \Sigma^0 + \frac{1}{2\sqrt{3}}\Lambda^0 &  \frac{1}{\sqrt{2}} n \\
	\frac{1}{\sqrt{2}} \Xi^- & \frac{1}{\sqrt{2}} \Xi^0 & - \frac{1}{\sqrt{3}} \Lambda^0
	\end{array} \right),
\end{align}
including the quark mass matrix as a perturbation to the $SU(3)$ flavor symmetry,
\begin{align}
	{\cal L}_{\rm mass} =& -2 M_0 {\rm Tr} \bar{B} B
	- 2 F {\rm Tr}  \bar{B} [M, B] + 2D {\rm Tr} \bar{B}\{M, B\} \nonumber \\
	& - \Delta_{\rm EM} (\bar{p} p + \bar{\Sigma}^+ \Sigma^+ + \bar{\Sigma}^- \Sigma^- + \bar{\Xi}^- \Xi^-).
\end{align}
Here, $F$ and $D$ are rescaled by $m_s$ compared to the standard notation to be made dimensionless, and $\Delta_{\rm EM}$ is the electromagnetic correction. Using the PDG values of baryon masses and quark masses, we find $F=2.11$, $D=0.641$, $M_0=1192$~MeV, $\Delta_{\rm EM}=5.0$~MeV, reproducing the octet baryon masses within 0.5\% errors. Now replacing $m_u$ with $\kappa\varphi$, we find the coupling to the proton
\begin{align}
	{\cal L}_{aNN} =& 1.47 \kappa \varphi\, \overline{p_L} p_R + c.c,
	\label{eq:Yukawa}
\end{align}
while no coupling to the neutron. %If we allow for the Kaplan--Manohar ambiguity here, the coupling can be smaller by a factor of $R$. 

Even though the coupling is $O(1)$, it is not as large as other hadronic couplings $g_{\pi N N} \approx 13$, $g_{\rho N N} \approx 4$. On the other hand, $\sigma$ or $a$ are heavier than $\rho$ with mass $m_\rho \approx 763$~MeV and $\omega$ with $m_\omega = 783$~MeV. The impact of $\sigma$ or $a$ exchange would be buried in the hard-core potential due to the $\rho$- and $\omega$-exchange. 

\section{FCNC Constraints}

\begin{figure}[t]
\includegraphics[width=0.7\columnwidth]{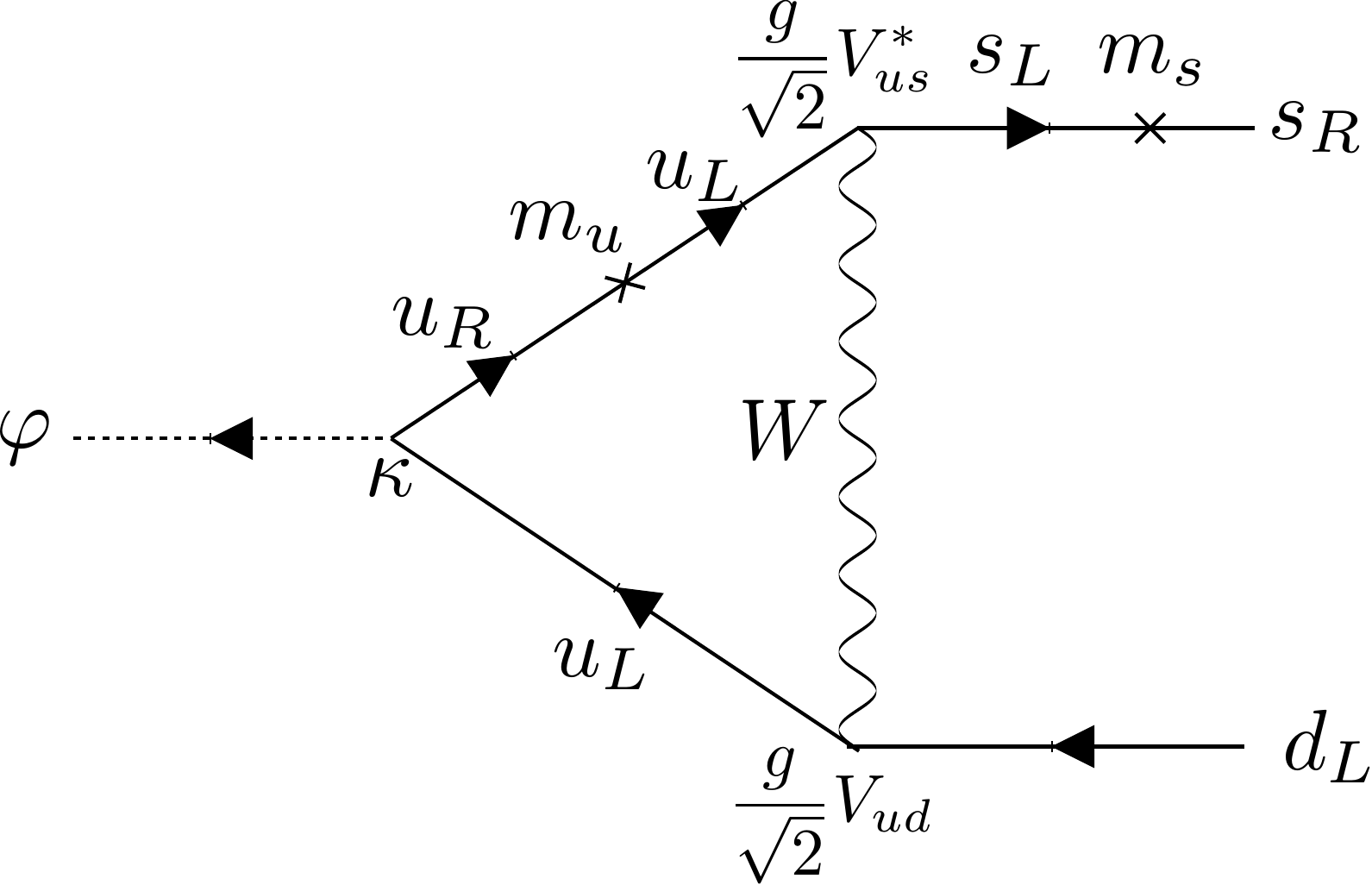}
\includegraphics[width=0.7\columnwidth]{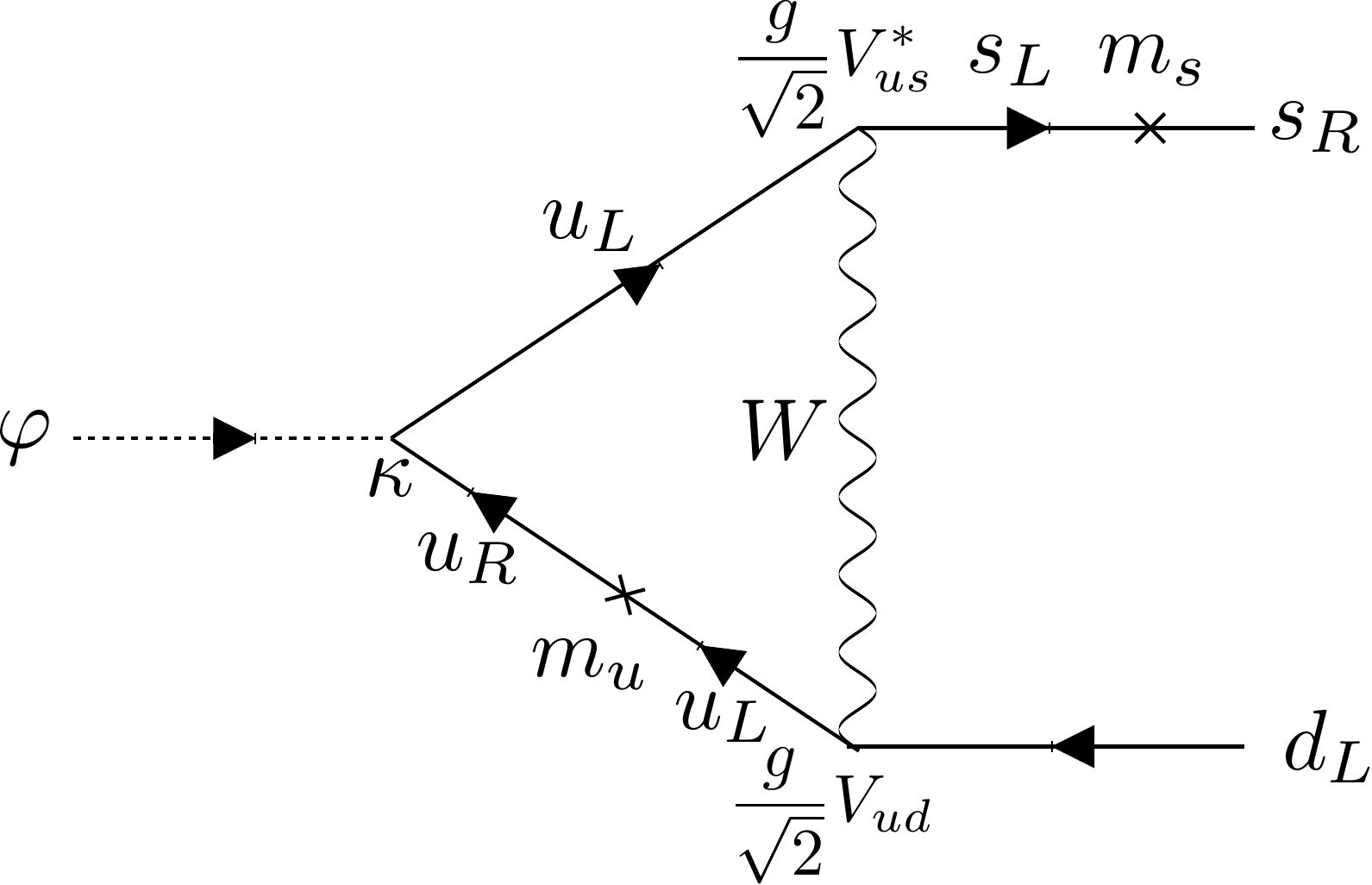}
\caption{The flavor-changing neutral-current processes by the scalar $\varphi$, incuding an operator Eq.~\eqref{eq:FCNC}.}
\label{fig:FCNC}
\end{figure}

Given that the $U(1)_{PQ}$ charge is flavor-dependent, the axion can induce flavor-changing neutral current (FCNC) processes as for variant axion \cite{Bardeen:1986yb} or flaxion \cite{Ema:2016ops}. 

The up-quark loop induces the coupling $\varphi^* \bar{s} d$ 
\begin{align}
	\frac{\kappa g^2}{16\pi^2} \sqrt{2}\, V_{us}^* V_{ud} 
	\frac{m_u m_s}{m_W^2} \left( \ln \frac{m_W^2}{m_u^2} -1 \right)\varphi^* \bar{s} d
	\label{eq:FCNC}
\end{align}
as well as $\varphi \bar{s} d$ (Fig.~\ref{fig:FCNC}).
It leads to $\bar{s} d \bar{u} u$ operator contributing to $K \rightarrow \pi \pi$ but it is one-loop suppressed relative to that of the standard model and further suppressed by $m_u m_s / m_\varphi^2$, hence negligible. The same comment applies to $b \rightarrow d \bar{u} u$. 

The $K^0$-$\overline{K}^0$ mixing is induced by using this operator twice. However the contribution is two-loop compared to the one-loop standard-model contribution and is further suppressed by $(m_u m_s)^2 / (m_\varphi^2 m_c^2)$, making it negligible. 

\section{Accelerator Constraints}

The paper \cite{Jaeckel:2015jla} studied LEP limits on the $a\gamma\gamma$ coupling using an effective operator $-\frac{1}{4} g_{a\gamma\gamma} a F_{\mu\nu} \tilde{F}^{\mu\nu}$, and excluded the mass range from MeV to 91~GeV down to $g_{a\gamma\gamma} \gtrsim 10^{-3}$--$10^{-2}$. Even though the description of this coupling using an effective operator from the triangle anomaly is valid for high $f_a$, it would be misleading for our model. In fact, there was an incorrect prediction that $Z \rightarrow \pi^0 \gamma$ branching fraction is as large as $10^{-3}$ using the triangle anomaly \cite{Jacob:1989pw}. It has been shown that the calculation based on the triangle anomaly is incorrect when photon is far off-shell (including on-shell $Z$) \cite{Deshpande:1990ej,Hikasa:1990xx,Manohar:1990hu,Pham:1990mf,West:1990ta,Young:1990pt,Schroder:1990nr,Chatterjee:1990eg,Maitra:1994pm,Micu:1995sc} (see Fig.~\ref{fig:agammagamma}). For our model, the process is suppressed by $g_{a\gamma\gamma} \sim \frac{\alpha}{2\pi} m_u/q^2$ and is negligible at LEP energies $q^2 \geq m_Z^2$. 

Since $\varphi$ does not have Yukawa couplings to charm or bottom quarks, limits on $J/\psi \rightarrow a\gamma$ or $\Upsilon \rightarrow a\gamma$ do not apply.  The coupling to the off-shell photon is suppressed with $q^2 = m_\Upsilon^2$ etc as in the case at LEP. 

Beam dump experiments such as \cite{Bross:1989mp} look for an axion that decays dominantly into $e^+ e^-$ or $\gamma\gamma$. The GlueX experiment looked for photoproduction of axion-like particle that decays into $\pi^0\pi^+\pi^-$ but the mass limit extends only up to 720~MeV \cite{GlueX:2021myx}.

\section{Hadroproduction}

Since the axion $a$ and its scalar partner $\sigma$ couple to protons Eq.~\eqref{eq:Yukawa}, they can be produced from hadronic collisions. Once produced, it is clear from the chiral Lagrangian \eqref{eq:chiral} that $\sigma$ decays to even number of pions or kaons, and $a$ to odd numbers of them. There are many scalar resonances that decay into $\pi\pi$, $f_0 (980)$, $f_0 (1370)$, $f_0 (1500)$, $f_0 (1710)$, $f_0 (1770)$, $f_0 (2020)$, $f_0 (2200)$, $f_0 (2470)$. It is possible that one of them is actually our $\sigma$ or its admixture with $q\bar{q}$, $qq\bar{q}\bar{q}$ or $gg$ configurations. Similarly, there are pseudoscalar resonances $\eta(1295)$, $\eta(1405)$, $\eta(1475)$, $\eta(1760)$, $\eta(2225)$, $\eta(2370)$ which may be our axion or its admixture with $q\bar{q}$ mesons.  There does not appear to be a degenerate pair of $f_0$ and $\eta$ but a splitting can be introduced with the quartic $(\varphi^*\varphi)^2$ term in the potential as mentioned earlier.  It is not easy to understand the entire resonance spectrum, especially when different configurations can mix with each other, and widths and branching fractions are poorly measured in many cases to pick up particular candidate states.

\section{Astrophysical Constraints}

The only astrophysical constraint for a heavy axion is from SN1987a, which applies to a mass up to 500~MeV \cite{lee2018revisitingsupernova1987alimits} if the dominant coupling is to photons. However, our axion couples to nucleons with the Yukawa coupling Eq.~\eqref{eq:Yukawa} and the axion is  trapped, not leading to cooling of the proto-neutron star core. We conclude that there are no astrophysical constraints.

\section{A UV Completion}

We need a dimension-five operator
\begin{align}
	\frac{y \kappa_u}{M_U} \bar{Q}_u \varphi H u_R + c.c.
	\label{eq:D5}
\end{align}
for the coupling $\kappa$ in Eq.~\eqref{eq:QCD}, with $\langle H \rangle = v=174$~GeV,
\begin{align} x
	\kappa &= \frac{yv}{M_{\cal Q}}\kappa_u  .
\end{align}

A UV completion of the operator Eq.~\eqref{eq:D5} has a heavy vector-like quark ${\cal Q}=({\cal U}, {\cal D})$ with the same quantum numbers as the left-handed quarks except for a $U(1)_{PQ}$ charge $-1$. The most general couplings are
\begin{align}
	{\cal L}_U =& \bar{\cal Q} (i{\not\!\! D} - M_{\cal Q}) {\cal Q} 
		+ (\kappa_i \bar{Q}_i \varphi {\cal Q} + y \bar{\cal Q} H u_R + c.c.)
\end{align}
Integrating out ${\cal Q}$, we indeed obtain Eq.~\eqref{eq:D5}. The vector-like quark can be produced in pairs at the LHC. The best limit $M_{\cal D}>1.53$~TeV comes from the pair production of ${\cal D}$ followed by ${\cal D} \rightarrow u_R W^-$ or its CP conjugate \cite{ATLAS:2024zlo}. With $y, \kappa_u \approx 1$ and $M_{\cal Q} \approx 1.6$~TeV, we find $\kappa \approx 0.11$, which is enhanced to $\kappa \approx 0.25$ down to the QCD scale, sufficiently large for our purposes. There is also a limit on ${\cal U}$ that decays into $u_L + \varphi$, which appear as four-jet events with di-jet resonances, limiting $M_{\cal U}> 770$~GeV \cite{CMS:2020zti} even though a few holes exist. A dedicated search for ${\cal U} \rightarrow u h$ would be very interesting. Single production searches at the LHC assume the $tW$ or $bW$ final states \cite{CMS:2021mku} and do not apply. For the light quark final states, the limits go back to the Tevatron experiments \cite{D0:2010ckq} that limit masses up to 700~GeV at most.

The light up-quark eigenstates have a mixing between ${\cal U}_L$ and $u_L$ only by $\kappa_u \langle \varphi \rangle / M_{\cal Q} \simeq 10^{-3}$. Therefore there is practically no modifications to the charged-current weak interactions and the impact on the CKM unitarity is only at the level of $10^{-6}$. On the other hand, the neutral-current weak interactions are affected by the large mixing between ${\cal U}_R$ and $u_R$, and hence the $Z$ coupling to the light right-handed up-quark eigenstate is modified as
\begin{align}
	-\frac{2}{3}\sin^2 \theta_W \rightarrow
	-\frac{2}{3}\sin^2 \theta_W + \frac{1}{2} \frac{(y v)^2}{M_{\cal Q}^2 + (yv)^2}\ .
\end{align}
For $M_{\cal Q}=1.6$~TeV and $y=1$, it modifies ${\rm BR}(Z \rightarrow u_R \overline{u_R})$ by $-0.15\%$, well within the measurement error of $\Delta {\rm BR}(Z \rightarrow u\bar{u}+c\bar{c}) = 0.6~\%$. On the other hand, the impact on the hadronic width is nominally at the $3\sigma$ level compared to the measurement error, $\Delta {\rm BR}(Z \rightarrow {\rm hadrons}) = 0.06~\%$. It will surely be tested by future Giga-$Z$ or Tera-$Z$ experiments. It is also subject to a question of contribution to the $Z$ decay into down-type quarks from new physics, beyond the scope of this work. 

The dimension-five operator \eqref{eq:D5} leads to a decay 
\begin{align}
	\lefteqn{
	\Gamma(h \rightarrow \overline{u_L} u_R \varphi, \overline{u_R} u_L \varphi^*)
	} \nonumber \\
	&= \left( \frac{y \kappa_u}{M_U} \right)^2 \frac{m_h^3}{512\pi^3} 
	= 0.21~{\rm MeV}  \left( \frac{y \kappa_u 770~{\rm GeV}}{M_U} \right)^2.
\end{align}
independent of the UV completion. It is comparable with the predicted $\Gamma ( h \rightarrow g g) = 0.335$~MeV \cite{ParticleDataGroup:2024cfk}. At the LHC, this decay mode would not be observable. At a future $e^+ e^-$ Higgs factory, the measurement of this decay mode can be probed as a part of the $\Gamma ( h \rightarrow {\rm hadrons})$ measurement with the bottom and charm veto.

The coupling $\kappa_c$ can lead to $D^0$--$\overline{D}^0$ mixing at the one-loop level with a box diagram of ${\cal U}$ and $\varphi$. This coupling is not needed in our proposal and we simply assume it is small. 

\section{Conclusion}

It is rather surprising that the Peccei--Quinn breaking scale may be below the QCD scale. In this case, the axion and its scalar partner are at the GeV range, and it is tantalizing to think that they may be hiding among the observed hadronic resonances. Given the high mass, it is immune to quantum gravity corrections. In a UV completed model, dijet resonance in four-jet events can be searched for at the LHC, and an enhanced $\Gamma ( h \rightarrow g g)$ at a future $e^+ e^-$ Higgs factory. The mass splitting between $\pi^\pm$ and $\pi^0$ is the most significant constraint.

\section{Acknowledgments}
\acknowledgments

HM thanks Tsutomu Yanagida, Ben Safdi, and Gilad Perez for useful comments. Special thanks go to Satoshi Shirai who checked aspects of UV completion. This work was supported by the NSF grant PHY-2515115, by the Director, Office of Science, Office of High Energy Physics of the U.S. Department of Energy under the Contract No.~DE-AC02-05CH11231, by the JSPS Grant-in-Aid for Scientific Research JP23K03382, Hamamatsu Photonics, K.K, Tokyo Dome Corporation, and by the World Premier International Research Center Initiative (WPI) MEXT, Japan.

\bibliography{refs}

\end{document}